\newcommand{\bear}{\begin{array}}  \newcommand{\eear}{\end{array}}
\newcommand{\bea}{\begin{eqnarray}}  \newcommand{\eea}{\end{eqnarray}}
\newcommand{\beq}{\begin{equation}}  \newcommand{\eeq}{\end{equation}}
\newcommand{\bef}{\begin{figure}}  \newcommand{\eef}{\end{figure}}
\newcommand{\bec}{\begin{center}}  \newcommand{\eec}{\end{center}}

\newcommand{\diracslash}[1]{#1\llap{/\kern2pt}}

\def\bearr{\begin{eqnarray}}
\def\eearr{\end{eqnarray}}

\documentclass[preprint,secnumarabic,floats,amssymb,nobibnotes,nofootinbib,aps,prc,showpacs,tightenlines]{revtex4}
\usepackage{shrthnds}
\usepackage{amssymb}
\usepackage{graphicx}
\usepackage{amsmath}
\usepackage{latexsym}
\usepackage{epsfig}
\usepackage{subfigure}
\usepackage[dvipsnames]{xcolor}
\begin{document}

\title{ Cosmological 
reconstruction  of   $f(T,\mathcal{T})$ Gravity}
\author{
Davood Momeni$^{}$\footnote{email: d.momeni@yahoo.com},
Ratbay Myrzakulov$^{}$\footnote{email: rmyrzakulov@gmail.com}}
\affiliation{Eurasian International Center for Theoretical Physics,\\
Eurasian National University, Astana 010008, Kazakhstan}
\date{\today}

\vspace{1.5 cm}
\begin{abstract}
\vspace{1.5 cm}
Motivated by the newly proposal for gravity as the effect of the torsion scalar $T$ and trace of the energy momentum tensor $\mathcal{T}$,we investigate the cosmological reconstruction of different models of the Universe. Our aim here is to show that how this modified gravity model,  $f(T,\mathcal{T})$ is able to reproduce different epoches of the cosmological history. We explicitly show that  $f(T,\mathcal{T})$  can be reconstructed for  
$\Lambda $CDM as the most popular and consistent model. Also we study the mathematical reconstruction of  $f(T,\mathcal{T})$  for a flat cosmological background filled by two fluids mixture. Such model describes  phantom-non-phantom era as well as the purely phantom
cosmology. We extend our investigation to more cosmological models like perfect fluid,Chaplygin gas and massless  scalar field. In each case we obtain 
some specific forms of $f(T,\mathcal{T})$. These families of  $f(T,\mathcal{T})$  contain arbitrary function of torsion and trace of the energy momentum.
\end{abstract}


\maketitle
\section{Introduction}
Now we have a real challenge to find the best model of gravity as the fundamental force of the world. Gravity is modelled by gauge theories. Classical gauges prepare a frame for quantum theory. Such model of gravity must be working for two major purposes:\par
1-Cosmology: Different obervational data 
 as type Ia supernovae , cosmic microwave background (CMB) , large scale structure , baryon acoustic oscillations , and weak lensing  indicate we live in an accelerating Universe \cite{data}. We know that the classical model of gravity,namely, General Relativity (GR) is not able to solve or describe this acceleration in a good and reasonable form.\par
2- High Energy Physics: Need to have a quantum theory of gravity and beyond on it, need to find a unified model for all fundamental forces of the Universe as an old wish of scientist. Such theory is needed now more than other times due to the recent progresses on experimental high energy physics.\par
or both of these reasons and to resolve these problems at the fundamental level, we need some modifications to the Einstein-Hilbert action (for a review of modified gravities see \cite{review}). Different modification has been proposed during decades. The first and the simplest one is to replace the Ricci scalar "R" with an arbitrary function , namely f (R), proposed originally in \cite{Buchdahl:1983zz} and motivated by different authors (for review see\cite{f(R)}). Another modification is when we insert $G$ the invariant of Gauss-Bonnet (GB) $G=R^2-4R_{\mu\nu}R^{\mu\nu}+ R_{\mu\nu \lambda\sigma}R^{\mu\nu\lambda\sigma}$. Also a unified description by $R, G$ exists and has several interesting properties\cite{GB}. \textbf{Different GB corrected models has been introduced to explain the dynamic of dark energy and to solve the problem of the currently accelerating expansion of the Universe
\cite{Nojiri:2005jg,Cognola:2006eg}. These "stringy inspired" models have been shown that are able to explain the acceleration expansion of the Universe. Indeed,the theory which we explained in this paper is motivated by the GB corrected models  \cite{Nojiri:2005jg,Cognola:2006eg}. }
\par
Historically meanwhile another description of the gravity as gauge theory existed in which gravity is the effect of the torsion not curvature. This model is called 
Teleparallel Theory and since at the level of the gravitational action and equations of the motion is equivalent to the GR, is called as Teleparallel Theory Equivalent (TEGR) (for a review see\cite{TEGR}). The original TEGR is not conformally invariant, but a version has been proposed to preserve this symmetry\cite{maluf} \par
Recently inspired from f (R), new scenario proposed as gravity in the form of  $f(T)$, where here $T$ is torsion scalar. It achieves many activities from different points of view \cite{f(T)}. A proposal for gravity is we take it as the effect of the curvature and the matter fields, namely the model as $f(R,\mathcal{L}_m,T_{\mu}^{(m)\nu},R^{\mu\nu}T^{(m)}_{\mu\nu})$\cite{coupled}. Such models exist and are consistent with the Mach's principle. One of such extended models is $f(R,\mathcal{T})$,here$\mathcal{T}$ denotes the trace of the energy momentum tensor \cite{f(RT)}. The model has several interesting features from thermodynamic to dynamics \footnote{Also, in \cite{Kiani:2013pba} , a similar scenario has been investigated.}.
\par
Now it was the time to unify $f(T)$ in the form of a non minimally coupled model like $f(R,\mathcal{T})$ and introduce it as a valid modified gravity. Very recently in \cite{Harko:2014aja},the authors proposed a new model of gravity $f(T,\mathcal{T})$ in which $f$ is an arbitrary function of torsion scalar and the trace of the energy momentum tensor. In this work we investigate cosmological reconstruction in the framework of this newly proposed for gravity. Before reconstruction technique proposed in \cite{sergei2009} and was used in  $f(T)$ \cite{Reconstruction}. So, for $f(T,\mathcal{T})$ we have enough motivation to study this problem.\par
The paper is organized as follows: in Section II we review the basic features of $f(T)$ gravity as the building block of the $f(T,\mathcal{T})$. In Section III we review $f(T,\mathcal{T})$. In Section IV, we present the general framework of reconstruction technique for $f(T,\mathcal{T})$. Sections V- XI are then devoted to construction of different cosmological epochs. We conclude the paper in last Section.

\section{f(T) gravity:
Quick review}
Since we deal with to an extension of  $f(T)$ theory,so it is needed we firstly review the program of $f(T)$. We start by writing the physical Riemannian metric of theory in basis of tetrads one-forms $\theta^i$:
\begin{eqnarray}
ds^2=g_{\mu\nu}dx^\mu dx^\nu=\eta_{ij}\theta^i\theta^j\,,
\end{eqnarray}
We define the coordinate representation of $\theta^i$ as the following: 
\begin{eqnarray}
d^\mu=e_{i}^{\;\;\mu}\theta^{i}; \,\quad \theta^{i}=e^{i}_{\;\;\mu}dx^{\mu}.
\end{eqnarray}
We denote by  $\eta_{ij}=diag(1,-1,-1,-1)$. Its just the Minkowskian metric . We also define $\{e^{i}_{\;\mu}\}$ by:
\begin{eqnarray}
e^{\;\;\mu}_{i}e^{i}_{\;\;\nu}=\delta^{\mu}_{\nu},\quad e^{\;\;i}_{\mu}e^{\mu}_{\;\;j}=\delta^{i}_{j}.
\end{eqnarray}
 In Weizenbo"ck's spacetimes the basic object to define the geometry is an analogy of the Levi-Civita's connection is defined by:
\begin{eqnarray}
\Gamma^{\lambda}_{\mu\nu}=e^{\;\;\lambda}_{i}\partial_{\mu}e^{i}_{\;\;\nu}=-e^{i}_{\;\;\mu}\partial_\nu e_{i}^{\;\;\lambda}.
\end{eqnarray}
We need to construct the objects like Ricci tensor, Riemann and others in this new geometry. So as the first attempt we define:
\begin{eqnarray}
T^{\lambda}_{\;\;\;\mu\nu}= \Gamma^{\lambda}_{\mu\nu}-\Gamma^{\lambda}_{\nu\mu},
\end{eqnarray}
The second  object is called as contorsion tensor :
\begin{eqnarray}
K^{\mu\nu}_{\;\;\;\;\lambda}=-\frac{1}{2}\left(T^{\mu\nu}_{\;\;\;\lambda}-T^{\nu\mu}_{\;\;\;\;\lambda}+T^{\;\;\;\nu\mu}_{\lambda}\right)\,\,.
\end{eqnarray}
Finally the useful tensor is $S_{\lambda}^{\;\;\mu\nu}$ :
\begin{eqnarray}
S_{\lambda}^{\;\;\mu\nu}=\frac{1}{2}\left(K^{\mu\nu}_{\;\;\;\;\lambda}+\delta^{\mu}_{\lambda}T^{\alpha\nu}_{\;\;\;\;\alpha}-\delta^{\nu}_{\lambda}T^{\alpha\mu}_{\;\;\;\;\alpha}\right)\,\,.
\end{eqnarray}
Now we are ready to define a quantity (scalar) in analogous to the Ricci scalar , we called it the torsion scalar :
\begin{eqnarray}
T=T^{\lambda}_{\;\;\;\mu\nu}S^{\;\;\;\mu\nu}_{\lambda}\,.
\end{eqnarray}
The general gravitational action for $f (T) $ in analogous to the $f (R) $ gravity is written as:
\begin{eqnarray}
 S= \int e \left[\frac{f(T)}{2\kappa^2} +\mathcal{L}_{m} \right]d^{4}x   \label{eq9}\,,
\end{eqnarray}
Here as GR, $\kappa^{2} = 8 \pi G $.\par
To find the equation of the motion (EOM) us perform a variation of the action (\ref{eq9}) with respect to the basis tetrads:
\begin{eqnarray}
S^{\;\;\; \nu \rho}_{\mu} \partial_{\rho} T f_{TT} + 
[e^{-1} e^{i}_{\;\; \mu}\partial_{\rho}(e e^{\;\; \mu}_{i}S^{\;\;\; \nu\lambda}_{\alpha} )
+T^{\alpha}_{\;\;\; \lambda \mu}   S^{\;\;\; \nu \lambda}_{\alpha} ]f_{T}+
\frac{1}{4}\delta^{\nu}_{\mu}f=\frac{\kappa^{2}}{2} \mathcal{T}^{\nu}_{\mu}  \label{eq10}\,,
\end{eqnarray}
Here $\mathcal{T}^{\nu}_{\mu}$ is the energy 
momentum tensor of the matter Lagrangian $\mathcal{L}_{m} $, $f_{T} = df(T)/dT$ and 
$f_{TT}  = d^{2}f(T)/dT^{2}$.
\par
To investigate cosmological reconstruction we adopt a flat $FRW$ cosmological background with the following common form:
\begin{eqnarray}
 ds^{2}= dt^{2} - a^{2}(t)\left(dx^2+dy^2+dz^2\right)\,.  \label{eq10'}
 \end{eqnarray}
A suitable and simple basis for tetrads is  the following diaginal one :
\begin{eqnarray}
 \{e^{a}_{\;\; \mu}\}= diag[1,a,a,a]. \label{eq11}
\end{eqnarray}
For this metric by computing the associated tensors, we obtain the torsion scalar :
\begin{eqnarray}
 T= -6H^{2}, \label{m1}
\end{eqnarray}
We define the Hubble parameter$H=\dot{a}/a$ . We suppose that our Universe is filled with a perfect fluid with the simple barotropic equation of state (EoS) $p_ {m} = w_ {m} \rho_{m}$. The associated energy momentum tensor is given by:
\begin{eqnarray}
\mathcal{T}^{\nu}_{\mu} =
diag(\rho_m,p_{m}, p_{m},
p_{m} ) . 
\end{eqnarray}
So, finally EOM or as we know,
modified Friedmann equations are presented by:
\begin{eqnarray}
 -Tf_T+\frac{1}{2}f&=& \kappa^{2} \rho_{m} \,, \\
 2\dot{T}Hf_{TT }+2\left(\dot{H}+3H^2\right)f_T+\frac{1}{2}f&=&-\kappa^2 p_m.
\end{eqnarray}
What we presented in this section is the starting point for the reconstruction of cosmological models in $f(T)$ and later $f(T,\mathcal{T})$ in the next sections.

\section{ $f(T,\mathcal{T})$ Gravity}
Very recently an extension of f(T) introduced \cite{Harko:2014aja}.\textbf{ This modified $f(T)$ gravity is inspired by the GB corrected models as it has been investigated in literature  \cite{Nojiri:2005jg,Cognola:2006eg}.} In this section we review the basics of $f(T,\mathcal{T})$ gravity. Let us  start by writing
 the gravitational  action:
\begin{equation}
S= \frac{1}{16\,\pi\,G}\,\int d^{4}x\,e\,\left[T+f(T,\mathcal{T})\right]%
+\int d^{4}x\,e\,\mathcal{L}_{m},
\label{action1}
\end{equation}%
This is an extension of $f(R,\mathcal{T})$ in which the matter field is coupled non minimally to the geometry by $\mathcal{T}$. It locally violates the equivalence principle. In this model $f(T,\mathcal{T})$ denotes an arbitrary algebraic function which it constructed from the torsion scalar $T$ and meanwhile by the traces $\mathcal{T}$ .
For simplicity we assume that the matter Lagrangian density $\mathcal{L}_{m}$ is only a well posed function of tetrads while it has no contribution from the derivatives of tetrads, it means $\frac{\partial \mathcal{L}_{m}}{\partial (\partial_{i}e_{a}^{\mu})}=0$ .\par
We perform the varietion of (\ref{action1}) like f (T), to find the modified EOM:
\begin{eqnarray}
&&\!\!\!\!\!\!\!\!\!\left(1+f_{T}\right) \left[e^{-1} \partial_{\mu}{(e
e^{\alpha}_{A}
S_{\alpha}^{~\rho \mu})}-e^{\alpha}_{A} T^{\mu}_{~\nu \alpha} S_{\mu}^{~\nu
\rho}\right]  \notag \ \ +\left(f_{TT} \partial_{\mu}{T}+f_{T\mathcal{T}} \partial_{\mu}{%
\mathcal{T}}\right) e^{\alpha}_{A} S_{\alpha}^{~\rho \mu}+ e_{A}^{\rho}
\left(\frac{f+T}{4}\right)  \notag \\
&&\ \ -f_{\mathcal{T}} \left(\frac{e^{\alpha}_{A} \overset{\mathbf{em}}{T}%
{}_{\alpha}^{~~\rho}+p e^{\rho}_{A}}{2}\right)=4\pi G e^{\alpha}_{A}
\overset%
{\mathbf{em}}{T}_{\alpha}{}^{\rho}.
\end{eqnarray}
here 
\begin{eqnarray}
f_{\mathcal{T}}=\partial{f}/\partial{\mathcal{T}},\ \
f_{T\mathcal{T
}}=\partial^2{f}/\partial{T} \partial{\mathcal{T}}.
\end{eqnarray}
\par
With FRW and diagonal tetrads like previous section the modified Friedmann equations are written:
\begin{equation}
H^{2}=\frac{8\pi G}{3} \rho_m-\frac{1}{6}\left(f+12 H^{2} f_{T}\right)+f_{%
\mathcal{T}} \left(\frac{\rho_{m}+p_{m}}{3}\right),
\label{Friedmann1}
\end{equation}
The second equation is obtained as:
\begin{eqnarray}
\dot{H}=-4\pi G \left(\rho_m+p_m\right)-\dot{H} \left(f_{T}-12 H^{2}
f_{TT}\right)  \notag \\
\ \ \ \ \ \ \, -H \left(\dot{\rho}_{m}-3\,\dot{p}_{m}\right)
f_{T\mathcal{T%
}}-f_{\mathcal{T}} \left(\frac{\rho_{m}+p_{m}}{2} \right).
\end{eqnarray}
It is adequate to write effective energy density and pressure for a pair of the field equations as the following:
\begin{equation}
\rho _{eff}=-\frac{1}{16\pi G}\left[ f+12f_{T}H^{2}-2f_{\mathcal{T}}\left(
\rho _{m}+p_{m}\right) \right].
\end{equation}%
\begin{eqnarray}
&&p_{eff}=: \left( \rho _{m}+p_{m}\right)\times  \notag \\
&&\hspace{-0.275cm}  \left[ \frac{1+f_{\mathcal{T}}/8\pi G}{1+f_{T}-12H^{2}f_{TT}+H\left( d\rho
_{m}/dH\right) \left( 1-3c_{s}^{2}\right) f_{T\mathcal{T}}}-1\right]
\notag \\
&& \hspace{-0.275cm}  +\frac{1}{16\pi G}\left[ f+12H^{2}f_{T}-2f_{\mathcal{T}}\left( \rho
_{m}+p_{m}\right) \right] .
\end{eqnarray}%
So, consequently we are able to rewrite  the modified Friedmann equations in the following commonly used forms:
\begin{eqnarray}
H^{2} &=&\frac{\kappa^2}{3}\rho_{tot}. \\
\dot{H} &=&-\frac{\kappa^2}{2}(\rho_{tot}+p_{tot}) .
\end{eqnarray}
Where
\begin{eqnarray}
&&\rho_{tot}= \rho _{DE}+\rho _{m}\\
&&p_{tot}=p_{DE}+p_{m} .
\end{eqnarray}
The continuty equation for $\rho_{tot},p_{tot}$ is:
\begin{equation}
\dot{\rho} _{tot}+3H\left(\rho _{tot}+p_{tot}\right)=0.
\end{equation}
\par
In the simple case of a perfect fluid, with a barotropic fluid, $\mathcal{T}=\rho_{m}-3\,p_{m}$. After this brief presentation of the model we investigate the cosmological reconstruction.
\section{A cosmological reconstruction technique in $f(T,\mathcal{T})$:
General framework}
The reconstruction 
method is a useful method to realize cosmological models ,has been proposed in \cite{Nojiri:2006gh},\cite{Nojiri:2006be}. 
Following the methodology of reconstruction \cite{sergei2009},for any cosmological era we have an equation in which Hubble parameter is a function of the scale factor:
\begin{eqnarray}
&&H^2=H^2(H^2_0,\rho_i^{0},a(t)).
\end{eqnarray} 
Where $\rho_i^{0}$ denotes the current values of different matter densities. Using this fundamental expression, we can write (\ref{Friedmann1}) as the following first order partial differential equation for $f=f(T,\mathcal{T})$:
\begin{eqnarray}
&& h(T,\mathcal{T})-\frac{1}{6}\left(f-2T f_{T}\right)+ g(T,\mathcal{T})f_{
\mathcal{T}}=0\label{eqf}.
\end{eqnarray}
Where,
\begin{eqnarray}
&&h(T,\mathcal{T})=\frac{T}{6}+\frac{8\pi G}{3} \rho_m,\\
&&g(T,\mathcal{T})=\left(\frac{\rho_{m}+p_{m}}{3}\right)
\end{eqnarray}
It has a family of solutions for a given set of functions $\{h(T,\mathcal{T}),g(T,\mathcal{T})\}$.
\section{$\Lambda$CDM model in  $f(T,\mathcal{T})$}
The first simple but important example is a model of the Universe in which we have matter fields with density $\rho_m$ and a positive cosmological constant $\Lambda$. The model is called as $\Lambda$CDM. The basic equation of $\Lambda$CDM era is given by:
\begin{eqnarray}
H^2=H_0^2+\frac{\kappa^2 }{3}\rho_0 a^{-3}
\end{eqnarray}
We adopt the units $\kappa^2=8\pi ,c=1$. Using the form of the energy momentum tensor, we obtain:
\begin{eqnarray}
\mathcal{T}=\rho_m,\ \ p_m=0.
\end{eqnarray}
The solution for (\ref{eqf}) has a very simple form:
\begin{eqnarray}
&&f(T,\mathcal{T})=-6H_0^2+\sqrt{|\mathcal{T}|}F(\frac{\mathcal{T}}{T})\label{LCDM}.
\end{eqnarray}
The first term on the right side is a torsion scalar at the present time, the second term contains an arbitrary function $F$. So, a wide class of the cosmological models exists. The viability of this family is an important issue to be addressed via data. If we choice $f(T,\mathcal{T})$ as (\ref{LCDM}),   $\Lambda$CDM model emerges.

\section{$f(T,\mathcal{T})$  reproducing the system with phantom and
non-phantom matter}
Second example on reconstruction deals with an era of the Universe in which the Universe is filled with a two component fluid mixture. One is phantom, with density $\rho_p a^ {c_1} $ and the second one is non-phantom with the density of matter field $\rho_q a^ {-c_1} $ \cite{sergei2009}. Since both densities satisfy continuity equations separately, the barotropic parameters $\{w_p,w_q\}$ are given by:
\begin{eqnarray}
w_p=-\frac{c_1}{3}-1,\ \ w_q=\frac{c_1}{3}-1.
\end{eqnarray}
When $w_q>-1$ (early Universe), the dominant epoch is non-phantom and in the present epoch (late time) $w_p-1$, so we live in the acceleration era and phantom is the dominant portion.\par
The fundamental equation is:
\begin{eqnarray}
H^2=\frac{\kappa^2}{3}(\rho_q a^{-c_1}+\rho_p a^{c_1})
\end{eqnarray}
If we set $c_1=4$ we obtain:
\begin{eqnarray}
&&\mathcal{T}=\rho_m-3p_m=(1-3w_{q})\rho_q a^{-4}+(1-3w_p)\rho_{p}a^4=8\rho_p a^4.
\end{eqnarray}
So,exact solution of (\ref{eqf}) reads:
\begin{eqnarray}
&&f(T,\mathcal{T})=\sqrt{|T|}G\Big(\frac{\sqrt[3]{|T|}(-64\rho_p+\rho_q \mathcal{T}^2)}{\rho_q}\Big).
\end{eqnarray}
Here $G$ is an arbitrary function.
\section{de-Sitter Universe in $f(T,\mathcal{T})$  Gravity}
In the early Universe and also at the late time (asymptotically) the solution of the scale factor is exponential, $a (t) =a_0 e^ {H_0 t} $, with a constant
Hubble parameter . The model is called de- Sitter, and the model of the Universe is the de- Sitter Universe. This is an exact solution of gravitational field equations with dust, the pressureless perfect fluid . So, we obtain:
\begin{eqnarray}
H=H_{0},\ \ \rho_m=\frac{3}{\kappa^2}H_0^2,\ \ p_m=0.
\end{eqnarray}
Now exact solution for (\ref{eqf}) is given by:
\begin{eqnarray}
&&f(T,\mathcal{T})=\sqrt{|T|}H(\mathcal{T}).
\end{eqnarray}
Where $H$ is an arbitary algebraic function.
\section{Einstein Static Universe in $f(T,\mathcal{T})$ Model}
Static scenario of the Universe, as an exact solution of Einstein field equations was the first solution. The solution is vaccum, no source of any kinds. It defines vanishing Hubble $H=0,\mathcal{T}=0$, consequently $T=0$. Exact solution of (\ref{eqf}) is:
\begin{equation}
f(T,\mathcal{T})=0.
\end{equation}
There is no Einstein Universe in  $f(T,\mathcal{T})$  gravity. We assume that $\lim_{\mathcal{T}\to 0} f_{\mathcal{T}}<\infty$.
\section{Reconstruction of perfect fluids in $f(T,\mathcal{T})$}
Perfect fluids with barotropic EoS 
$p=\omega\rho$ is a good (simple) example (no physically viable) for scenarios of accelerated expansion. For this source field, we calculate:
\begin{eqnarray}
&&\tau=(1-3\omega)\rho.
\end{eqnarray}
So, (\ref{eqf}) reads:
\begin{eqnarray}
&&-\frac{T}{6}=\frac{\kappa^2}{3(1-3\omega)}\mathcal{T}-\frac{1}{6}(f-2Tf_T)+\frac{(1+\omega)}{3(1-3\omega)}\mathcal{T}f_{\mathcal{T}}.
\end{eqnarray}
Exact solution for this equation is:
\begin{eqnarray}
f(T,\mathcal{T})=\sqrt{|T|}U\Big(\mathcal{T}T^{\frac{1+\omega}{3\omega-1}}\Big)-T+\frac{4\kappa^2(1-\omega)}{(3\omega-1)(5\omega+1)}\mathcal{T}T^{\frac{1+\omega}{3\omega-1}+\frac{2(1-\omega)}{5\omega+1}}.
\end{eqnarray}
Here $U$ is an arbitrary function.

\section{Reconstruction Using Chaplygin Gas}
One of the most interesting models for accelerated expansion is the Chaplygin gas (CG). The model inspired from fluids in aerodynamic. The case of CG is defined by:
\begin{eqnarray}
p=\frac{-A}{\rho}.
\end{eqnarray}
 This is the simplest example of CG. Generalized models are also amazing models but we are investigating just this simple model. Using it, we find:
\begin{eqnarray}
&&\mathcal{T}=\rho+\frac{3A}{\rho},\ \ \rho(\mathcal{T})=\frac{\mathcal{T}\pm\sqrt{\mathcal{T}^2-12A}}{2}.
\end{eqnarray}
Explicitly exact solution for (\ref{eqf}) reads:
\begin{eqnarray}
&&f(T,\mathcal{T})=\frac{e^{-\frac{B(\mathcal{T})}{2}}}{2}\Big(2V\Big(Te^{B(\mathcal{T})}\Big)+\int_{0}^{\mathcal{T}}
{\frac{\rho(x)dx}{A-\rho(x)^2}}\Big[Te^{B(\mathcal{T})-\frac{1}{2}B(x)}+2\kappa^2\rho(x)e^{\frac{1}{2}B(x)}\Big]\Big).
\end{eqnarray}
Where:
\begin{eqnarray}
&&B(x)=\int_{0}^{x}{\frac{\rho(y)dy}{A-\rho(y)^2}}.
\end{eqnarray}
The model has so complicated form, here $V$ is an arbitrary function.
\section{$f(T,\mathcal{T})$ Models for Scalar Field}
As the last instructive example we study the cosmological reconstruction for the massless scalar field. Such simple fields are important in high energy physics due to the invariance under conformal transformations and solvability. The simple form of a minimally coupled scalar field is given by:
\begin{eqnarray}
\mathcal{L}_m=-\frac{1}{2}\omega\phi_{,\mu}\phi^{,\mu}.
\end{eqnarray}
Where $\omega$ is a dimensionless parameter. Energy momentum tensor for this matter Larangian is:
\begin{eqnarray}
T_{\mu\nu}=-\frac{1}{2}\omega\Big(g_{\mu\nu}(\phi_{,\alpha}\phi^{,\alpha})-2\phi_{,\mu}\phi_{,\nu}\Big).
\end{eqnarray}
Using this tensor,trace reads :
\begin{eqnarray}
&&\rho_m=T_{0}^{0}=\frac{1}{2}\omega\dot{\phi}^2,\\
&&p_m=-T_{1}^{1}=\frac{1}{2}\omega\dot{\phi}^2=\rho_m,\\
&&\mathcal{T}=-2\rho_m.
\end{eqnarray}
Or:
\begin{eqnarray}
&&p_m=\rho_m=-\frac{\mathcal{T}}{2}.
\end{eqnarray}

Exact solution for (\ref{eqf}) is given by:
\begin{eqnarray}
&&f(T,\mathcal{T})=-\frac{2\kappa^2}{3}\mathcal{T}-T+\sqrt{|T|}W(\mathcal{T}T).
\end{eqnarray}
The first term is TEGR, the second term is proportional to the Hubble parameter and the last term is an arbitrary function $W$.

 \section{Conclusion}
 $f(T,\mathcal{T})$ is an extension of $f (T) $ gravity in which $\mathcal {T} $ is trace of the energy momentum tensor of matter fields. It's a healthy extension in f (T) gravity. Cosmological reconstruction is a powerful technique in modified gravity invented to find the analogy of different cosmological epochs in modified gravity. In this paper we investigated cosmological reconstruction in $f(T,\mathcal{T})$. In any epoch we have a fundamental equation of Hubble and EoS. Using this information we find $f(T,\mathcal{T})$. Such reconstructed models are able to explain the current accelerating expansion of the Universe in the context of modified gravities. For this purpose we must investigate effective as of dark energy. For a class of simple models it is already done \cite{Harko:2014aja}.

{\bf Acknowledgement}: We would like to thank  Emmanuel N. Saridakis for useful comments and the anonymous reviewer for enlightening comments related to this work. 


\begin{thebibliography}{17}

\addcontentsline{toc}{chapter}{Bibliographie}

\bibitem{data}
S. Perlmutter et al. [SNCP Collaboration], Astrophys. J. {\bf 517}, 565 (1999); A. G. Riess et al.[SNST Collaboration], Astron. J. {\bf 116}, 1009 (1998);
D. N. Spergel et al. [WMAP Collaboration], Astrophys. J. Suppl. {\bf 148}, 175 (2003); ibid. 170,
377 (2007); E. Komatsu et al. [WMAP Collaboration], ibid. {\bf 180}, 330 (2009);
E. Komatsu et al. [WMAP Collaboration], Astrophys. J. Suppl. {\bf 192}, 18 (2011);
M. Tegmark et al., Phys. Rev. D {\bf 69}, 103501 (2004); U. Seljak et al. [SDSS Collaboration],
Phys. Rev. D {\bf 71}, 103515 (2005);
D. J. Eisenstein et al., Astrophys. J. {\bf 633}, 560 (2005);
B. Jain and A. Taylor, Phys. Rev. Lett. {\bf 91}, 141302 (2003).
\bibitem{review}
S. Capozziello and M. De Laurentis, Phys. Rept. 509,
167 (2011); 
E. J. Copeland, M. Sami and S. Tsujikawa, Int. J. Mod. Phys. D 15, 1753 (2006);S.~'i.~Nojiri and S.~D.~Odintsov,
  eConf C {\bf 0602061}, 06 (2006)
  [Int.\ J.\ Geom.\ Meth.\ Mod.\ Phys.\  {\bf 4}, 115 (2007)]
  [hep-th/0601213].
\bibitem{Buchdahl:1983zz} 
  H.~A.~Buchdahl,
  Mon.\ Not.\ Roy.\ Astron.\ Soc.\  {\bf 150}, 1 (1970).

\bibitem{f(R)}
A.~De Felice and S.~Tsujikawa,
  Living Rev.\ Rel.\  {\bf 13} (2010) 3
  [arXiv:1002.4928 [gr-qc]];
S.~'i.~Nojiri and S.~D.~Odintsov,
  Phys.\ Rept.\  {\bf 505} (2011) 59
  [arXiv:1011.0544 [gr-qc]].


\bibitem{GB}
G. Cognola, E. Elizade, S. Nojiri, S. D. Odintsov and S. Zerbini, 
Phys. Rev. D \textbf{73} 084007 (2006) [arxiv:hep-th/0601008];
  E.~Elizalde, R.~Myrzakulov, V.~V.~Obukhov and D.~S\'aez-G\'omez,
  Class. Quant. Grav. \textbf{27} (2010) 095007
  [arXiv:1001.3636 [gr-qc]];
A.~De Felice, J.~-M.~Gerard and T.~Suyama,
Phys.\ Rev.\ D {\bf 82}, 063526 (2010)
\textbf{\bibitem{Nojiri:2005jg} 
  S.~'i.~Nojiri and S.~D.~Odintsov,
  Phys.\ Lett.\ B {\bf 631}, 1 (2005)
  [hep-th/0508049].}
\textbf{\bibitem{Cognola:2006eg} 
  G.~Cognola, E.~Elizalde, S.~'i.~Nojiri, S.~D.~Odintsov and S.~Zerbini,
  Phys.\ Rev.\ D {\bf 73}, 084007 (2006)
  [hep-th/0601008].}





\bibitem{TEGR}
F. W. Hehl, J. D. McCrea, E. W. Mielke and Y. Ne’eman, Phys. Rep. 258, 1 (1995).
\bibitem{maluf}
J. W. Maluf and F. F. Faria, Annalen Phys. 524, 366 (2012) [arXiv:1203.0040 [gr-qc]]..
\bibitem{f(T)}
R. Ferraro and F. Fiorini, Phys. Rev. D 75, 084031 (2007); G. R. Bengochea,  R. Ferraro, Phys. Rev. D, 79, 124019, (2009).; E. V. Linder, Phys. Rev. D 81, 127301 (2010);S. Capozziello, V. F. Cardone, H. Farajollahi and A. Ravanpak, Phys. Rev. D 84, 043527 (2011);K. Bamba, S. Capozziello, M. De Laurentis, S. ’i. Nojiri and D. Sez-Gmez, Phys. Lett. B 727, 194 (2013); S. Basilakos, S. Capozziello, M. De Laurentis, A. Paliathanasis and M. Tsamparlis, Phys. Rev. D 88, 103526 (2013);  A. Paliathanasis, S. Basilakos, E. N. Sari- dakis, S. Capozziello, K. Atazadeh, F. Darabi and M. Tsamparlis, arXiv:1402.5935 [gr-qc]; S. Capozziello, P. A. Gonzalez, E. N. Saridakis and Y. Vasquez, JHEP 1302 (2013) 039.
\bibitem{coupled}
T. Harko, Phys. Lett. B 669, 376 (2008); T. Harko and F. S. N. Lobo, Eur. Phys. J. C 70, 373 (2010); T. Harko, F. S. N. Lobo and O. Minazzoli, Phys. Rev. D 87, 047501 (2013); J. Wang and K. Liao, Class. Quant. Grav. 29, 215016 (2012);S. Capozziello, G. Lambiase and H. J. Schmidt, Annalen Phys. 9, 39 (2000).

\bibitem{Kiani:2013pba} 
  F.~Kiani and K.~Nozari,
  Phys.\ Lett.\ B {\bf 728}, 554 (2014)
  [arXiv:1309.1948 [gr-qc]].
\bibitem{f(RT)}
T. Harko, F. S. N. Lobo, S. Nojiri and S. D. Odintsov, “f(R, T ) gravity,” Phys. Rev. D {\bf 84} (2011) 024020. [arXiv:1104.2669 [gr-qc]];
  M.~Jamil, D.~Momeni, M.~Raza and R.~Myrzakulov,
  Eur.\ Phys.\ J.\ C {\bf 72}, 1999 (2012)
  [arXiv:1107.5807 [physics.gen-ph]];
  M.~Jamil, D.~Momeni and R.~Myrzakulov,
  Chin.\ Phys.\ Lett.\  {\bf 29} (2012) 109801
  [arXiv:1209.2916 [physics.gen-ph]];M. J. S. Houndjo, Int. J. Mod. Phys. D. {\bf 21}, 1250003 (2012). arXiv: 1107.3887 [astro-ph.CO]; M.~J.~S.~Houndjo,
  Int.\ J.\ Mod.\ Phys.\ D {\bf 21} (2012) 1250003
  [arXiv:1107.3887 [astro-ph.CO]];H.~Shabani and M.~Farhoudi,
  Phys.\ Rev.\ D {\bf 88}, 044048 (2013)
  [arXiv:1306.3164 [gr-qc]].
\bibitem{Harko:2014aja}
  T.~Harko, F.~S.~N.~Lobo, G.~Otalora and E.~N.~Saridakis,
  arXiv:1405.0519 [gr-qc].
\textbf{\bibitem{Nojiri:2006gh} 
  S.~'i.~Nojiri and S.~D.~Odintsov,
  Phys.\ Rev.\ D {\bf 74}, 086005 (2006)
  [hep-th/0608008].
\bibitem{Nojiri:2006be} 
  S.~'i.~Nojiri and S.~D.~Odintsov,
  J.\ Phys.\ Conf.\ Ser.\  {\bf 66}, 012005 (2007)
  [hep-th/0611071].}





\bibitem{sergei2009}  S.~'i.~Nojiri, S.~D.~Odintsov and D.~Saez-Gomez,
  Phys.\ Lett.\ B {\bf 681}, 74 (2009)
  [arXiv:0908.1269 [hep-th]].

\bibitem{Reconstruction}
 I.~G.~Salako, M.~E.~Rodrigues, A.~V.~Kpadonou, M.~J.~S.~Houndjo and J.~Tossa,
  JCAP {\bf 1311}, 060 (2013)
  [arXiv:1307.0730 [gr-qc]];M.~Jamil, D.~Momeni and R.~Myrzakulov,
  Eur.\ Phys.\ J.\ C {\bf 72}, 2137 (2012)
  [arXiv:1210.0001 [physics.gen-ph]];
  M.~U.~Farooq, M.~Jamil, D.~Momeni and R.~Myrzakulov,
  Can.\ J.\ Phys.\  {\bf 91}, 703 (2013)
  [arXiv:1306.1637 [astro-ph.CO]];M.~Hamani Daouda, M.~E.~Rodrigues and M.~J.~S.~Houndjo,
  Eur.\ Phys.\ J.\ C {\bf 72}, 1893 (2012)
  [arXiv:1111.6575 [gr-qc]].




\end{thebibliography}
\end{document}